\documentclass[onecolumn]{aastex62}

\usepackage{bm}
\usepackage{soul}
\usepackage{multirow}
\usepackage{lineno}
\graphicspath{./}

\begin{document}

\title{V606 Cen: a newly formed massive contact binary in a hierarchical triple system}

\correspondingauthor{Li Fu-xing}
\email{lfx@ynao.ac.cn}

\author{Li F.-X.}
\affiliation{Yunnan Observatories, Chinese Academy of Sciences(CAS), P.O. Box 110, 650216 Kunming, P.R. China}
\affiliation{University of Chinese Academy of Sciences, Yuquan Road 19\#, Sijingshang Block, 100049 Beijing, China}
\affiliation{Key Laboratory of the Structure and Evolution of Celestial Objects, CAS, Kunming 650216, P.R. China}

\author{Liao W.-P.}
\affiliation{Yunnan Observatories, Chinese Academy of Sciences(CAS), P.O. Box 110, 650216 Kunming, P.R. China}
\affiliation{Key Laboratory of the Structure and Evolution of Celestial Objects, CAS, Kunming 650216, P.R. China}

\author{Qian S.-B.}
\affiliation{Yunnan Observatories, Chinese Academy of Sciences(CAS), P.O. Box 110, 650216 Kunming, P.R. China}
\affiliation{University of Chinese Academy of Sciences, Yuquan Road 19\#, Sijingshang Block, 100049 Beijing, China}
\affiliation{Key Laboratory of the Structure and Evolution of Celestial Objects, CAS, Kunming 650216, P.R. China}

\author{Fern{\'a}ndez Laj{\'u}s E.}
\affiliation{Facultad de Ciencias Astron{\'o}micas y Geof\'isicas, Universidad Nacional de La Plata, Paseo del Bosque s/n,\\
       1900 La Plata, Buenos Aires, Argentina}
\affiliation{Instituto de Astrof\'isica de La Plata (CCT La Plata $-$ CONICET/UNLP), 1900 La Plata, Argentina}

\author{Zhang J.}
\affiliation{Yunnan Observatories, Chinese Academy of Sciences(CAS), P.O. Box 110, 650216 Kunming, P.R. China}
\affiliation{Key Laboratory of the Structure and Evolution of Celestial Objects, CAS, Kunming 650216, P.R. China}

\author{Zhao E.-G.}
\affiliation{Yunnan Observatories, Chinese Academy of Sciences(CAS), P.O. Box 110, 650216 Kunming, P.R. China}
\affiliation{Key Laboratory of the Structure and Evolution of Celestial Objects, CAS, Kunming 650216, P.R. China}

\begin{abstract}
V606 Centauri (V606 Cen) is an early B-type close binary with an orbital period of 1.4950935\,d and the complete light curves are very difficult to be observed on the ground. By analyzing the continuous light curve obtained by TESS, we found that it is a marginal contact binary with a very low fill-out factor of about 2\%. The O-C diagram of V606 Cen is constructed for the first time based on the 118.8-years of eclipse times. The O-C diagram has been found to show a downward parabolic change together with a cyclic oscillation with a semi-amplitude of 0.0545\, d and a period of 88.3\, yr.
The downward parabolic variation reveals a linear period decrease at a rate of $dP/dt = -2.08 \times{10^{-7}} d \cdot yr^{-1}$ that can be explained by the mass transfer from the more massive component to the less massive one. Both the marginal contact configuration and the continuous period decrease suggest that V606 Cen is a newly formed contact binary via Case A mass transfer. The cyclic change in the O-C diagram can be explained by the Light-Travel Time Effect via the presence of a third body. The lowest mass of the tertiary companion is determined as M$_{3}$ = 4.51($\pm0.43$)M$_{\odot}$ and the tertiary is orbiting around the central eclipsing binary in a nearly circular orbit (e=0.33). All the results indicate that V606 Cen is a newly formed massive contact binary in a hierarchical triple system.

\end{abstract}


\section{Introduction} \label{sec:intro}
\label{sect:intro}

\label{sect:intro}
Massive stars play an important role in the properties of galaxies (Massey 2003). Among the early OB-type stars, binary stars are key in exploring evolutionary process. More significantly, there is a very high fraction of binaries in OB-type stars. For example, 56\% of O-type objects are probable close binaries, and over 70\% of all massive stars will exchange mass with a companion (Sana et al. 2012). The presence of a nearby component will alter the evolution (de Mink et al. 2013). Therefore, we can study evolutionary models by analyzing the evolution stages in many early-type binary stars (Lorenzo et al. 2014; Qian et al. 2014; Polushina 2004). The massive binaries may evolve into a contact configuration via a fast/slow Case A or Case B mass transfer (Sybesma 1985, 1986; Pols 1994), and many of these contact binaries will merge (Langer 2012) by common envelope evolution or form to be compact objects and core-collapse supernovas and so on. Eclipsing binaries are important in the determination of the fundamental, and it can determine the fundamental stellar parameters from photometry and spectroscopy.

V606 Cen is an early B-type massive eclipsing binary, and it was found to be a variable star by Swope (1939). It was classified as a $\beta$ Lyr-type light curve with the period 1.495108 \,d, and the first ephemeris was determined (Hertzsprung 1950). The spectral type was first obtained in 1971 (Stephenson \& Sanduleak), and the outcome was $OB^-$. Four years later, its spectral type was revised as B1-2 Ib/II by Houk (1975). It was first noted as a contact binary by Eggen (1978). The light curve was presented by Mayer et al. (1992) who observed three minima. V606 Cen was studied in detail for the first time by Lorenz et al. (1999), and no authors have studied it again after that. They presented high-resolution CCD spectroscopy and the photometric light curve in the Johnson $UBV$ bands. On the one hand, the radial velocity curve and the spectral mass ratio $q_{spec}$ were determined. On the other hand, the photometric solution was accomplished based on the MORO code, and the contact configuration for V606 Cen was confirmed. The absolute parameters of the two components were also derived. When considering the effect of radiation pressure on early spectral type over-contact binaries, Bauer (2005) modeled V606 Cen with radiation pressure. The results were in good agreement with that of Lorenz et al. (1999) calculation and corresponded well to the predictions based on detached binaries. We now have the opportunity to further study the evolution state of V606 Cen with better data and better methods, such as the high-precision continuous photometry data of the Transiting Exoplanet Survey Satellite (TESS, Ricker et al. 2015) and the Digital Access to a Sky Century at Harvard (DASCH, Grindlay et al. 2012a, 2012b) database that provides about 100 years of a photometric times series. More information about the derived minima is described in Section 2.

For the evolution stage of V606 Cen, Lorenz et al. (1999) made the initial discussion without consideration of the orbital. They proposed that V606 Cen has probably evolved into a contact binary during the slow phase of case A mass transfer. We collected minima spanning over 118.8-years, and the O-C diagram was constructed and analyzed for the first time. Firstly, the existence of the additional body was confirmed by the Light-Travel Time Effect. Secondly, we can estimate the parameters of the additional body.
The formation of early massive contact binary stars is still unsolved. At present there are relatively few known OB-type massive contact binaries (Table 1). Moreover, the results and evolution stage of V606 Cen are different from from these cases (e.g., evolutionary state, mass transfer, geometrical structure, etc.).
The explanations of the columns in Table 1:
Column 1 is stars name.
Column 2 is period of binaries.
Column 3 is mass ratio $\mathnormal{q} = M_{less}/M_{more}$.
Column 4 is the mass of more massive components.
Column 5 is the mass of less massive components.
Column 6 is degree of contact factor.
Column 7 is structure type of O-C.
Column 8 is period of third body.
Column 9 is lowest mass of third body.
Column 10 is rates of periodic change of binaries.
Column 11 is rates of the mass transfer.
Column 12 is spectral types.
Column 13 is references.
In this paper, we present a new analysis of the light curve that demonstrates that V606 Cen is actually a triple system observed at a special point in its evolutionary history.

\begin{table}
\caption{The samples of the early OB-type contact binaries}
\begin{center}
\begin{tabular}{llllllccccccc}
\hline\hline
Name  &Period &q                &$M_{1}$     &$M_{2}$     &f  &O$-$C &$P_{3}$ &$M_{3}$          &dP$/$dt            &dM$/$dt         &Spec.  &Ref.\\
      &(d)   &$M_{2}$/$M_{1}$  &($M_{\odot}$) &($M_{\odot}$) &\% &Type  &(yr)   &min($M_{\odot}$) &$\times{10}^{-7}$  &$\times{10}^{-7}$  &  &\\
\hline
CT Tau	&0.667 	&0.983	&14.25	&14.01	&99	    &b	&64.6	&1.29	&\nodata	&\nodata	        &B1V+B1V  &1\\
V593 Cen &0.755 &0.952	&6.2	&5.9	&45.4	&b	&50.9	&4.3	&\nodata	&\nodata	&B5+B5  &2\\
V701 Sco &0.762 &0.996	&9.78	&9.74	&55.4	&a	&41.2	&1.86	&+0.00118	&\nodata	&B1+B1.5  &1,3,4,5\\
BH Cen	&0.792 	&0.89	&9.4	&7.9	&46.4	&a	&50.3	&2.2	&+1.26	&28	    &B3+B3  &3,6\\
BR Mus	&0.798 	&0.96	&\nodata	    &\nodata	&11	&\nodata	&\nodata	&\nodata	&\nodata	&\nodata	&B3  &7\\
GU Mon	&0.897 	&0.976	&8.79	&8.58	&72.4	&c	&34.2	&1.89	&-5.09	&\nodata	    &B1V+B1V  &1,8\\
MY Cam	&1.175 	&0.84	&37.7	&31.6	&\nodata	&\nodata	&\nodata	&\nodata	&\nodata	&\nodata	&O5.5V+O6.5V  &9\\
TU Mus	&1.387 	&0.652	&16.7	&10.4	&\nodata	    &a	&47.73	&1.55	&+4.0	&4.2    &O7V$+$O8V  &10,11,12\\
V745 Cas &1.410 &0.572	&18.31	&10.47	&\nodata	&\nodata	&\nodata	&\nodata	&\nodata	&\nodata	&B0V+B1-2V  &13\\
SV Cen	&1.659 	&0.707	&8.56*	&6.05*	&79	&\nodata	&\nodata	&\nodata	&\nodata	&\nodata	&B1V+B6.5III  &14,15\\
V599 Aql &1.849 &0.672	&6.1	&4.1	&\nodata	&\nodata	&\nodata	&\nodata	&\nodata	&\nodata	&B3+B5  &16\\
V382 Cyg &1.886 &0.742	&26	    &19.3	&10    &a	&47.7	&2.56	&+4.4	&4.3	&O7.3V$+$O7.7V &10,17,18\\
LY Aur	&4.002 	&0.528	&30	    &18.6	&\nodata	    &a	&12.5	&3.4	&+7.2	&\nodata	    &O9V+O9.5V  &3,19\\
V729 Cyg &6.598 &0.303	&30.4	&9.2	&\nodata	&\nodata	&\nodata	&\nodata	&\nodata	&\nodata	&O7+O8  &20\\\hline
OGLE SMC &0.883 &0.845	&16.9	&14.7	&70	&\nodata	&\nodata	&\nodata	&\nodata	&\nodata	&O9+O9.5-B0  &18,21\\
SC10 108086 &	&	&	& &	&	&	& &	&	&	&\\
VFTS 352 &1.120 &0.98	&25.6	&25.1	&29	&\nodata	&\nodata	&\nodata	&\nodata	&\nodata	&O5.5V+O6.5V  &18,22\\
VFTS 066 &1.140 &0.508	&13	&6.6	&\nodata	&\nodata	&\nodata	&\nodata	&\nodata	&\nodata	&O9V+B0.2V  &22\\
OGLE-SMC &2.206 &\nodata	&\nodata	&\nodata	&\nodata	&b	&12.4	&\nodata	&\nodata	&\nodata	&O5+O6  &23\\
ECL-4690 &	&	&	& &	&	&	& &	&	&	&\\
\hline
\end{tabular}
\end{center}
Note: *M{$sin^{3}(i)$}; (a) Upward parabola trend superimposed cyclic variation; (b) Cyclic variation; (c) Downward parabola trend superimposed cyclic variation;\\
Reference: (1) \citet{2019AJ....157..111Y}; (2) \citet{2019ApJ...871L..10E}; (3) \citet{2014NewA...26..112Z}; (4) \citet{2006NewA...12..117Q}; (5) \citet{1987MNRAS.226..899B}; (6) \citet{2018RAA....18...59Z}; (7) \citet{1987RMxAA..14..402L}; (8) \citet{2021AJ....162...13L}; (9) \citet{2014A&A...572A.110L}; (10) \citet{2007MNRAS.380.1599Q}; (11) \citet{2008ApJ...681..554P}; (12) \citet{2021MNRAS.507.5013M}; (13) \citet{2014MNRAS.442.1560C}; (14) \citet{1992AJ....103..573R}; (15) \citet{2013CEAB...37...79Z}; (16) \citet{1991A&A...245..517H}; (17) \citet{1997MNRAS.285..277H}; (18) \citet{2021A&A...651A..96A}; (19) \citet{1994Obs...114..107S}; (20) \citet{2016Ap.....59...68A}; (21) \citet{2005MNRAS.357..304H}; (22) \citet{2020A&A...634A.119M}; (23) \citet{2017MNRAS.469.2952Z}.
\end{table}

\section{Data acquisition and new photometric observation}

The photometric data used in this paper for V606 Cen are based upon our observations and open available light curve databases.
Photometric observations were carried out on 2019 May 1, 4 and 2008 April 7 from the Complejo Astronomico El Leoncito (CASLEO), San Juan, Argentina, using a 0.60 m Helen Sawyer Hogg (HSH) telescope with a Johnson $V$-band filter. There are three eclipse minima shown in Fig. 1.
Another neglected minimum was found from the light curves in the Optical Monitoring Camera (OMC, Alfonso-Garz{\'o}n et al. 2012), as shown in the lower right panel of Fig. 1. The OMC is a small refractor telescope with a 5 cm diameter aperture and Johnson V-filter.
We also used the online open databases from the Transiting Exoplanet Survey Satellite (TESS, Ricker et al. 2015), the All Sky Automated Survey (ASAS, Pojmanski 2002), the Bochum Survey of the Southern Galactic Disk (GDS, Hackstein et al. 2015), and the Digital Access to a Sky Century at Harvard (DASCH, Grindlay et al. 2012a, 2012b).

TESS is a space telescope project. It is equipped with four cameras with a large field of view (24$\times$24 degree) and used to detect brighter stars. Each strip was monitored for about 27 days.
The orbital parameters were initially derived using the TESS data, which is a continuous and unbroken time-series photometry
data. Since the period of V606 Cen is close to 1.5 \,d, it is challenging to observe a complete phased light curve on the earth (e.g., Mayer et al. (2010) took more than two years to obtain the ground light curve from a total of 22 observation nights). However, the space TESS project can achieve a continuous light curve. For V606 Cen, we downloaded the images from TESSCut (Brasseur et al. 2019). There are two segments, tess-s0011-2-2 and tess-s0038-2-1, and their exposure time is 30 and 10 minutes, respectively. We extracted the light curve by applying the aperture photometry method to the candidate object and background (Dotson et al. 2019; Barentsen et al. 2019; Lightkurve Collaboration et al. 2018; Liu et al. 2021), which has proved to be reliable and trustworthy for eclipsing binaries. In the extraction, we set the percentile threshold of background and aperture to 15\% and 85\% of total pixels with a flux order, respectively, and the aperture radius was three pixels (Liu et al. 2021). Both extracted phase curves are equally good, and their mean errors are 0.0002 and 0.0003 mag. The results are shown in Fig.2.

We collected many eclipse minima from all the available databases and our observations. Two methods were used: parabola fitting directly for continuous observation data, and parabola fitting by reconstructing a phase with dispersed data of more than one cycle. The photographic plates of V606 Cen were digitized by the DASCH project, and they cover the years between 1889 and 1989. These data show large scatter and lower time resolution with a mean error of 0.32 mag. However, they are beneficial for orbital period analysis because of the long time span. Many authors have also used the second method (Shi et al. 2021; Zasche et al. 2017; Qian et al. 2021), and its reliability and accuracy have been demonstrated. There are 34 eclipse times from DASCH, eight eclipse times from ASAS and one eclipse time from GDS, that were determined by the method. In the ASAS database, only the data of the quality flag A and B were adopted. The samples from ASAS, DASCH and GDS are shown in Fig. 3. All the eclipse times are listed in Table 2.

\begin{figure}
\begin{center}
\includegraphics[angle=0,scale=0.45]{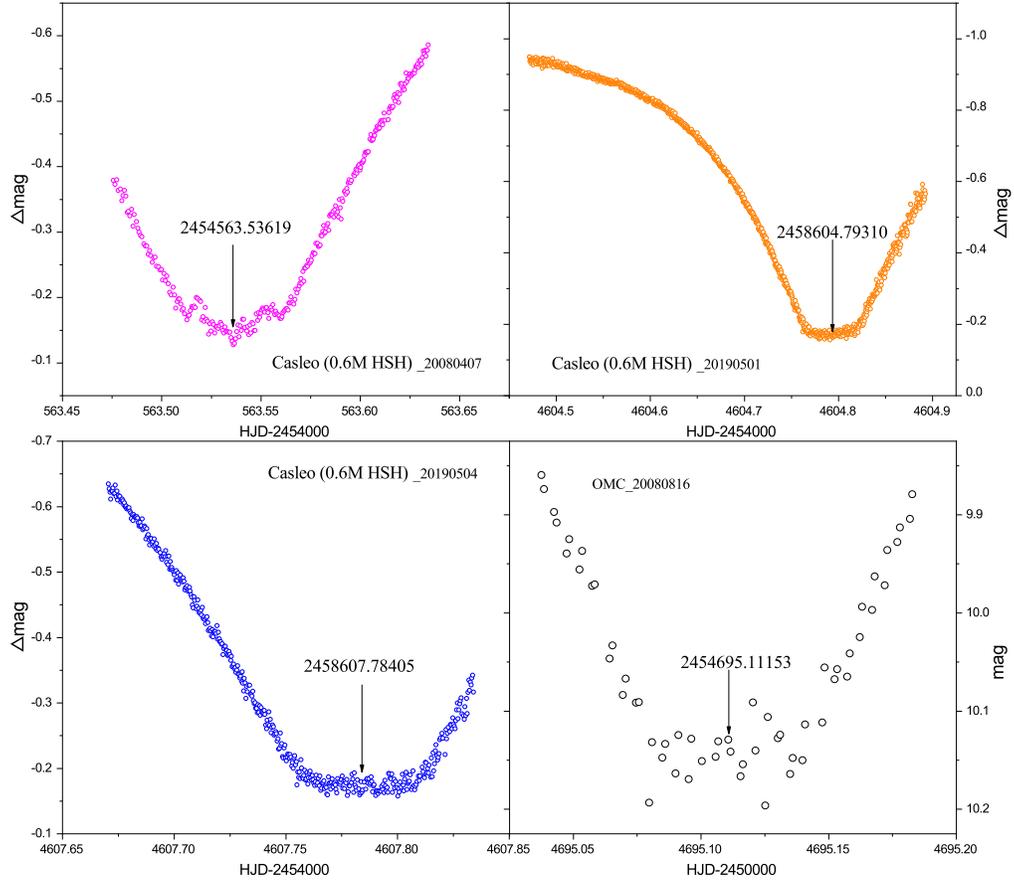}
\caption{The eclipse times by the CCD photometric observations from CASLEO and OMC.}
\end{center}
\end{figure}

\begin{figure}
\begin{center}
\includegraphics[angle=0,scale=0.45]{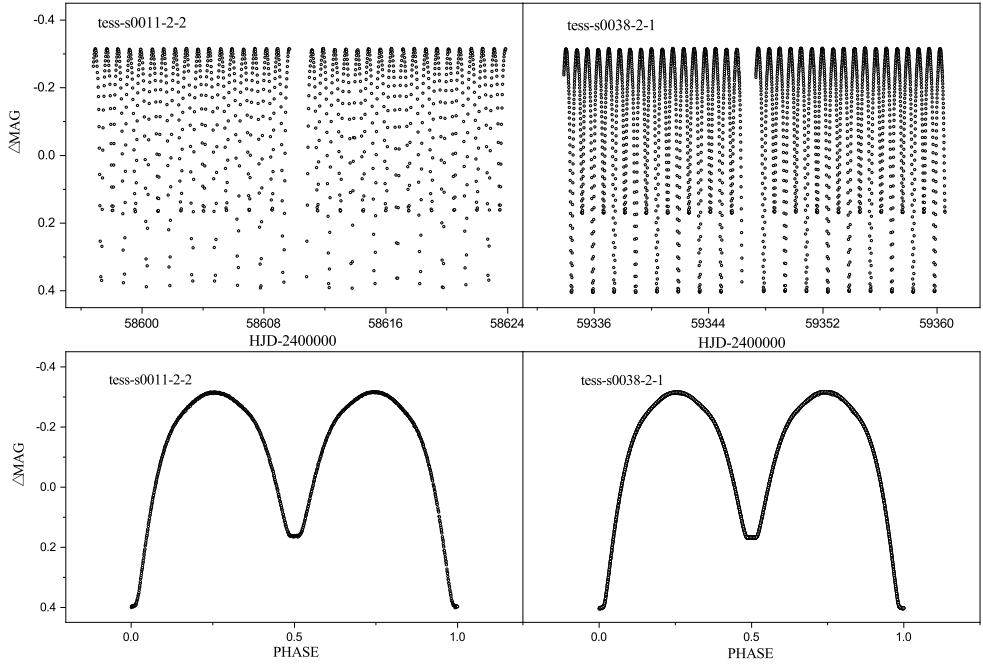}
\caption{The light curve and the phased curve of TESS data.}
\end{center}
\end{figure}

\begin{small}
\begin{longtable}{lllllllll}
\caption{\label{tab:} All the times of light minimum for V606 Cen}\\
\hline\hline
Eclipse-times  &Errors &Eclipse &Method &Source &Epoch &$O-C$  &Weight  &Reference\\
HJD-2400000    &($\pm d$)   &P/S     &       &          &      &(d) & &\\
\hline
\endfirsthead %1第一页表头

\caption{\label{tab:}(Continued)}\\
\hline\hline
Eclipse-times  &Errors &Eclipse &Method &Source &Epoch &$O-C$  &Weight  &Reference\\
HJD-2400000    &($\pm d$)   &P/S     &       &          &      &(d) & &\\
\hline \endhead %2续页表头

\hline
\multicolumn{6}{r}{\textsl{(Continued)}}\\
\endfoot

\hline
%\multicolumn{12}{r}{\textsl{Note. The one digital numbers in the parentheses are the errors on the last one bits of the data.}} \\
%\multicolumn{12}{r}{\textsl{The capital letters from A to U in the parentheses represent 21 multiplets.}} \\
%\multicolumn{12}{r}{\textsl{The asterisk indicates that the AMP-PH diagram trend of this frequency is different from the same multiplet.}} \\
%\multicolumn{6}{r}{\textsl{Here, $f0$ is the orbital frequency.}} \\

\endlastfoot
15959.27684 	&0.00763 	&S	&PG	&DASCH	&-8021.5	&-0.18465 	&1 	&1\\
16270.24973 	&0.00821 	&S	&PG	&DASCH	&-7813.5	&-0.19120 	&1 	&1\\
16544.61053 	&0.00756 	&P	&PG	&DASCH	&-7630	&-0.18007 	&1 	&1\\
16963.23875 	&0.00839 	&P	&PG	&DASCH	&-7350	&-0.17803 	&1 	&1\\
17461.10923 	&0.00992 	&P	&PG	&DASCH	&-7017	&-0.17368 	&1 	&1\\
18015.79964 	&0.00965 	&P	&PG	&DASCH	&-6646	&-0.16296 	&1 	&1\\
18645.23932 	&0.00852 	&P	&PG	&DASCH	&-6225	&-0.15764 	&1 	&1\\
18926.31490 	&0.00992 	&P	&PG	&DASCH	&-6037	&-0.15964 	&1 	&1\\
19548.28716 	&0.00852 	&P	&PG	&DASCH	&-5621	&-0.14628 	&1 	&1\\
20826.62548 	&0.00916 	&P	&PG	&DASCH	&-4766	&-0.11290 	&1 	&1\\
20976.12821 	&0.00835 	&P	&PG	&DASCH	&-4666	&-0.11952 	&1 	&1\\
21491.94110 	&0.00685 	&P	&PG	&DASCH	&-4321	&-0.11388 	&1 	&1\\
22271.66405 	&0.00818 	&S	&PG	&DASCH	&-3799.5	&-0.08219 	&1 	&1\\
23344.40151 	&0.00843 	&P	&PG	&DASCH	&-3082	&-0.07433 	&1 	&1\\
24990.51767 	&0.01088 	&P	&PG	&DASCH	&-1981	&-0.05611 	&1 	&1\\
27165.92788 	&0.01360 	&P	&PG	&DASCH	&-526	&-0.00694 	&1 	&1\\
27952.35400 	&\nodata    &P	&PG	&GCVS 	&0	&0.00000 	&1 	&4\\
28079.44022 	&0.01002 	&P	&PG	&DASCH	&85	&0.00327 	&1 	&1\\
28988.46130 	&0.00973 	&P	&PG	&DASCH	&693	&0.00751 	&1 	&1\\
29525.21128 	&0.00999 	&P	&PG	&DASCH	&1052	&0.01892 	&1 	&1\\
29857.12932 	&0.01195 	&P	&PG	&DASCH	&1274	&0.02621 	&1 	&1\\
30266.78011 	&0.01024 	&P	&PG	&DASCH	&1548	&0.02137 	&1 	&1\\
30796.04216 	&0.01459 	&P	&PG	&DASCH	&1902	&0.02032 	&1 	&1\\
31216.92583 	&0.00970 	&S	&PG	&DASCH	&2183.5	&0.03518 	&1 	&1\\
31702.09793 	&0.00963 	&P	&PG	&DASCH	&2508	&0.04943 	&1 	&1\\
32792.02629 	&0.00941 	&P	&PG	&DASCH	&3237	&0.05463 	&1 	&1\\
42912.28966 	&0.00974 	&P	&PG	&DASCH	&10006	&0.03010 	&1 	&1\\
43996.23569 	&0.00873 	&P	&PG	&DASCH	&10731	&0.03334 	&1 	&1\\
44159.18444 	&0.00807 	&P	&PG	&DASCH	&10840	&0.01690 	&1 	&1\\
44643.60631 	&0.01281 	&P	&PG	&DASCH	&11164	&0.02848 	&1 	&1\\
45213.22513 	&0.01516 	&P	&PG	&DASCH	&11545	&0.01667 	&1 	&1\\
45415.06169 	&0.00718 	&P	&PG	&DASCH	&11680	&0.01561 	&1 	&1\\

46020.55414 	&0.00817 	&P	&PG	&DASCH	&12085	&-0.00481 	&1 	&1\\
46132.67760 	&0.00941 	&P	&PG	&DASCH	&12160	&-0.01336 	&1 	&1\\
46341.99088 	&0.00840 	&P	&PG	&DASCH	&12300	&-0.01317 	&1 	&1\\
48684.81310 	&0.00040 	&P	&CCD	&ESO 0.5m	&13867	&-0.00246 	&10 	&2\\
48687.80230 	&0.00030 	&P	&CCD	&ESO 0.5m	&13869	&-0.00345 	&10 	&2\\
48690.79300 	&0.00030 	&P	&CCD	&ESO 0.5m	&13871	&-0.00294 	&10 	&2\\
49016.72400 	&0.00030 	&P	&CCD	&ESO 0.5m	&14089	&-0.00232 	&10 	&2\\
49019.71420 	&0.00040 	&P	&CCD	&ESO 0.5m	&14091	&-0.00231 	&10 	&2\\
49153.52610 	&0.00030 	&S	&CCD	&ESO 0.5m	&14180.5	&-0.00128 	&10 	&2\\
49457.77940 	&0.00020 	&P	&CCD	&ESO 0.5m	&14384	&0.00050 	&10 	&2\\
51968.04600 	&\nodata    &P	&CCD	&ASAS 	&16063	&0.00511 	&10 	&4\\
52415.08235 	&0.00147 	&P	&CCD	&ASAS	&16362	&0.00850 	&1 	&1\\
52715.59627 	&0.00203 	&P	&CCD	&ASAS	&16563	&0.00863 	&1 	&1\\
53124.5083      &0.0003     &S  &PE     &SAAO   &16836.5 &0.01259   &10 &3\\
53204.49226 	&0.00192 	&P	&CCD	&ASAS	&16890	&0.00905 	&1 	&1\\
53351.01126 	&0.00365 	&P	&CCD	&ASAS	&16988	&0.00888 	&1 	&1\\
53419.7860      &\nodata    &P  &CCD    &GCVS   &17034 	&0.00932    &1  &4\\
53575.27621 	&0.00223 	&P	&CCD	&ASAS	&17138	&0.00981 	&1 	&1\\
54153.87958 	&0.00398 	&P	&CCD	&ASAS	&17525	&0.01200 	&1 	&1\\
54331.79796 	&0.00401 	&P	&CCD	&ASAS	&17644	&0.01424 	&1 	&1\\
54563.53619 	&0.00053 	&P	&CCD	&CASLEO 0.6m	&17799	&0.01264 	&10 	&1\\
54695.11153 	&0.00118 	&P	&CCD	&OMC	&17887	&0.02009 	&10 	&1\\
54911.89675 	&0.00428 	&P	&CCD	&ASAS	&18032	&0.01675 	&1 	&1\\
56600.61013 	&0.00228 	&S	&CCD	&GDS	&19161.5	&0.02203 	&1 	&1\\
58597.28547 	&0.00248 	&P	&CCD	&TESS	&20497	&0.03173 	&1 	&1\\
58598.03302 	&0.00239 	&S	&CCD	&TESS	&20497.5	&0.03318 	&1 	&1\\
58598.78056 	&0.00267 	&P	&CCD	&TESS	&20498	&0.03263 	&1 	&1\\
58599.52811 	&0.00297 	&S	&CCD	&TESS	&20498.5	&0.03308 	&1 	&1\\
58600.27566 	&0.00321 	&P	&CCD	&TESS	&20499	&0.03154 	&1 	&1\\
58601.02320 	&0.00269 	&S	&CCD	&TESS	&20499.5	&0.03199 	&1 	&1\\
58601.77075 	&0.00223 	&P	&CCD	&TESS	&20500	&0.03044 	&1 	&1\\
58602.51830 	&0.00334 	&S	&CCD	&TESS	&20500.5	&0.03289 	&1 	&1\\
58603.26584 	&0.00273 	&P	&CCD	&TESS	&20501	&0.03335 	&1 	&1\\
58604.01339 	&0.00358 	&S	&CCD	&TESS	&20501.5	&0.03280 	&1 	&1\\
58604.76094 	&0.00268 	&P	&CCD	&TESS	&20502	&0.03225 	&1 	&1\\
58604.79310 	&0.00035 	&P	&CCD	&CASLEO 0.6m	&20502	&0.03216 	&10 	&1\\
58605.50848 	&0.00425 	&S	&CCD	&TESS	&20502.5	&0.03370 	&1 	&1\\
58606.25603 	&0.00213 	&P	&CCD	&TESS	&20503	&0.02915 	&1 	&1\\
58607.00358 	&0.00293 	&S	&CCD	&TESS	&20503.5	&0.03261 	&1 	&1\\
58607.75112 	&0.00221 	&P	&CCD	&TESS	&20504	&0.03206 	&1 	&1\\
58607.78405 	&0.00053 	&P	&CCD	&CASLEO 0.6m	&20504	&0.03293 	&10 	&1\\
58608.49867 	&0.00361 	&S	&CCD	&TESS	&20504.5	&0.03351 	&1 	&1\\
58609.24622 	&0.00267 	&P	&CCD	&TESS	&20505	&0.03096 	&1 	&1\\
58611.48886 	&0.00310 	&S	&CCD	&TESS	&20506.5	&0.03232 	&1 	&1\\
58612.23640 	&0.00219 	&P	&CCD	&TESS	&20507	&0.03077 	&1 	&1\\
58612.98395 	&0.00286 	&S	&CCD	&TESS	&20507.5	&0.03223 	&1 	&1\\

58613.73150 	&0.00196 	&P	&CCD	&TESS	&20508	&0.03168 	&1 	&1\\
58614.47904 	&0.00307 	&S	&CCD	&TESS	&20508.5	&0.03313 	&1 	&1\\
58615.22659 	&0.00237 	&P	&CCD	&TESS	&20509	&0.03359 	&1 	&1\\
58615.97414 	&0.00225 	&S	&CCD	&TESS	&20509.5	&0.03404 	&1 	&1\\
58616.72168 	&0.00163 	&P	&CCD	&TESS	&20510	&0.03149 	&1 	&1\\
58617.46923 	&0.00321 	&S	&CCD	&TESS	&20510.5	&0.03294 	&1 	&1\\
58618.21678 	&0.00226 	&P	&CCD	&TESS	&20511	&0.03040 	&1 	&1\\
58618.96433 	&0.00290 	&S	&CCD	&TESS	&20511.5	&0.03185 	&1 	&1\\
58619.71187 	&0.00202 	&P	&CCD	&TESS	&20512	&0.03230 	&1 	&1\\
58620.45942 	&0.00300 	&S	&CCD	&TESS	&20512.5	&0.03375 	&1 	&1\\
58621.20697 	&0.00230 	&P	&CCD	&TESS	&20513	&0.03321 	&1 	&1\\
58621.95451 	&0.00233 	&S	&CCD	&TESS	&20513.5	&0.03366 	&1 	&1\\
58622.70206 	&0.00274 	&P	&CCD	&TESS	&20514	&0.03211 	&1 	&1\\
58623.44961 	&0.00568 	&S	&CCD	&TESS	&20514.5	&0.03356 	&1 	&1\\
59334.40320 	&0.00041	&P	&CCD	&TESS	&20990	&0.03664 	&1 	&1\\
59335.15108 	&0.00046	&S	&CCD	&TESS	&20990.5	&0.03697 	&1 	&1\\
59335.89814 	&0.00043	&P	&CCD	&TESS	&20991	&0.03648 	&1 	&1\\
59336.64635 	&0.00043	&S	&CCD	&TESS	&20991.5	&0.03714 	&1 	&1\\
59337.39309 	&0.00046	&P	&CCD	&TESS	&20992	&0.03634 	&1 	&1\\
59338.14125 	&0.00045	&S	&CCD	&TESS	&20992.5	&0.03695 	&1 	&1\\
59338.88859 	&0.00041	&P	&CCD	&TESS	&20993	&0.03674 	&1 	&1\\
59339.63652 	&0.00043	&S	&CCD	&TESS	&20993.5	&0.03713 	&1 	&1\\
59340.38350 	&0.00042	&P	&CCD	&TESS	&20994	&0.03656 	&1 	&1\\
59341.13175 	&0.00045	&S	&CCD	&TESS	&20994.5	&0.03726 	&1 	&1\\
59341.87866 	&0.00041	&P	&CCD	&TESS	&20995	&0.03663 	&1 	&1\\
59342.62668 	&0.00042	&S	&CCD	&TESS	&20995.5	&0.03710 	&1 	&1\\
59343.37323 	&0.00049	&P	&CCD	&TESS	&20996	&0.03610 	&1 	&1\\
59344.12154 	&0.00047	&S	&CCD	&TESS	&20996.5	&0.03687 	&1 	&1\\
59344.86860 	&0.00042	&P	&CCD	&TESS	&20997	&0.03638 	&1 	&1\\
59345.61705 	&0.00043	&S	&CCD	&TESS	&20997.5	&0.03728 	&1 	&1\\
59347.85854 	&0.00046	&P	&CCD	&TESS	&20999	&0.03613 	&1 	&1\\
59348.60686 	&0.00044	&S	&CCD	&TESS	&20999.5	&0.03691 	&1 	&1\\
59349.35402 	&0.00041	&P	&CCD	&TESS	&21000	&0.03652 	&1 	&1\\
59350.10176 	&0.00047	&S	&CCD	&TESS	&21000.5	&0.03671 	&1 	&1\\
59350.84882 	&0.00043	&P	&CCD	&TESS	&21001	&0.03623 	&1 	&1\\
59351.59714 	&0.00043	&S	&CCD	&TESS	&21001.5	&0.03700 	&1 	&1\\
59352.34396 	&0.00041	&P	&CCD	&TESS	&21002	&0.03627 	&1 	&1\\
59353.09200 	&0.00043	&S	&CCD	&TESS	&21002.5	&0.03677 	&1 	&1\\
59354.58671 	&0.00046	&S	&CCD	&TESS	&21003.5	&0.03638 	&1 	&1\\
59355.33409 	&0.00041	&P	&CCD	&TESS	&21004	&0.03622 	&1 	&1\\
59356.08237 	&0.00046	&S	&CCD	&TESS	&21004.5	&0.03695 	&1 	&1\\
59356.82921 	&0.00043	&P	&CCD	&TESS	&21005	&0.03624 	&1 	&1\\
59357.57722 	&0.00045	&S	&CCD	&TESS	&21005.5	&0.03671 	&1 	&1\\
59358.32386 	&0.00046	&P	&CCD	&TESS	&21006	&0.03580 	&1 	&1\\
59359.07213 	&0.00047	&S	&CCD	&TESS	&21006.5	&0.03652 	&1 	&1\\
59359.81934 	&0.00041	&P	&CCD	&TESS	&21007	&0.03619 	&1 	&1\\
\hline
\end{longtable}
{\footnotesize References:} \footnotesize (1) This paper; (2) Lorenz et al. (1999); (3) Mayer et al. (2006); (4) O-C gateway \footnote{http://var2.astro.cz/ocgate/\\
}
\end{small}

\begin{figure}
\begin{center}
\includegraphics[angle=0,scale=0.4]{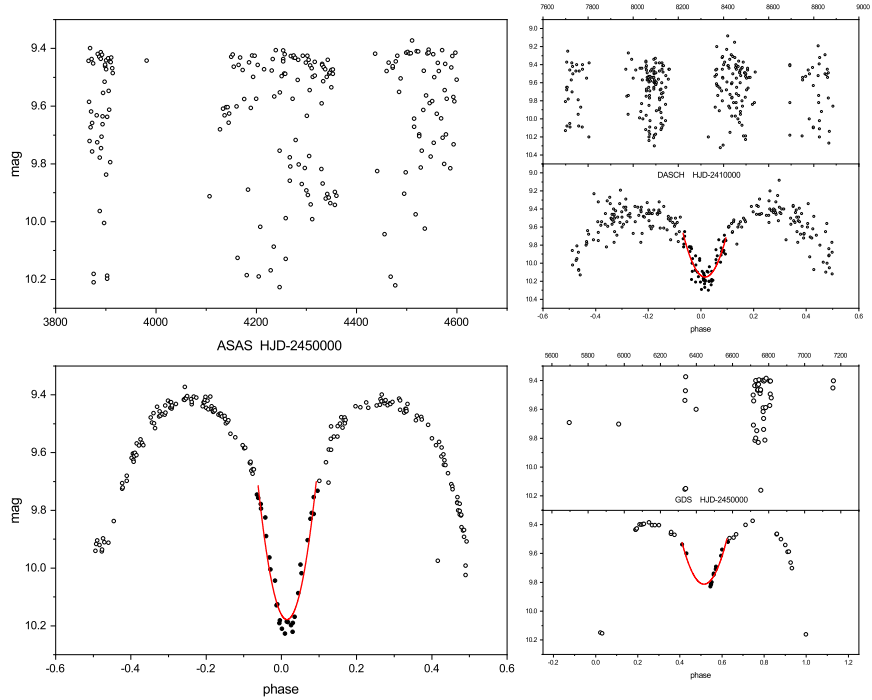}
\caption{Eclipse times from three databases (left for ASAS, upper right for DASCH, lower right for GDS). In each case, the upper panel show the times series and the lower panel shows the orbital phase light curve with the fitted parabola around the time of minimum.}
\end{center}
\end{figure}

\section{INVESTIGATION ON PERIOD VARIATIONS}

The analysis of periodic variations in the eclipse timings is a common way to study close binaries. It may help us to detect visible or invisible additional objects by the Light-Travel Time Effect and understand the mass transfer between their components by the parabolic trend.
V606 Cen is an early B-type contact binary, whose eclipse timings have not been studied before this work. We have analyzed its O-C diagram for the first time, and the data adopted span over 118.8-years. So far, there are few studies of the early OB-type contact binaries, which makes it difficult to confirm the evolution theory. We collected data on the known OB-type contact binaries in Table 1. The sample includes galactic (upper part) and extragalactic (lower part) binaries.
The O-C curve of V606 Cen with respect to the linear ephemeris was given from the O-C gateway (see earlier footnote):
\begin{equation}
Min.I = HJD\,2427952.354 + 1.4950935\times{E}.
\end{equation}

In the analysis of the O-C diagram, the eccentric orbit of the third celestial body rotating around the central binary was considered first. It was calculated based on equation (1), and the resulting O-C estimates are shown in the upper panel of Fig. 4. The O-C variations were fit based on the Light-Travel Time Effect by the following equations (Irwin 1952).
\begin{equation}
O-C=\Delta{T_0}+\Delta{P_0}\times{E}+{\beta}E^{2}+{\tau},
\end{equation}
\begin{equation}
{\tau}=K\frac{1}{(1-e^{2}\cos^{2}\omega)^{1/2}}[\frac{1-e^2}{1+e\cos\nu}\sin(\nu+\omega)+e\sin\omega].
\end{equation}
Where ${\tau}$, ${\beta}$ are the cyclic change and the rate of the linear period decrease. $\Delta{T_{0}}$ and $\Delta{P_{0}}$ in the formula are the revised epoch and period.
K is the semi-amplitude of O-C, e, $\nu$, $\omega$ are the eccentricity, the true anomaly and the longitude of the periastron from the ascending node for the additional body. The parameters in equation (2) and equation (3) were estimated using a weighted, least-squares fit with weights of 10 and 1, the former was assigned to photoelectric (PE) and CCD data from literature and observation except for TESS, because many eclipse-times of TESS were obtained in a very short time. The latter was applied to the photographic (PG), TESS and the reconstructing fitting minima.

The (O-C)$_{1}$ diagram shows a noticeable cyclic oscillation with a semi-amplitude of 0.0545\, d and a period of 88.3\, yr.
We tried to analyze the O-C diagram when the additional body with a circular orbit. Its (O-C)$_{1}$ also shows a clear periodic change with a small semi-amplitude in Fig.4. The residuals of both methods are shown in the lower panel of Fig. 4. The corresponding results of these two cases are listed in Table 3. From these two cases, their O-C diagrams appear as a downward parabola superimposed with a periodic change. The $\chi^{2}_{\nu}$ value of these two cases are 1.708 and 2.212, and the $\chi^{2}_{\nu}$ value of the eccentric orbit case is smaller. In order to detect the actual orbit parameters of the third body, more high precise minima data need to be observed, which may be obtained in the next cycle of about 20 years in order to record more than the 1.5 cycles of the Light-Travel Time Effect shown here.

\begin{figure}
\begin{center}
\includegraphics[angle=0,scale=0.4]{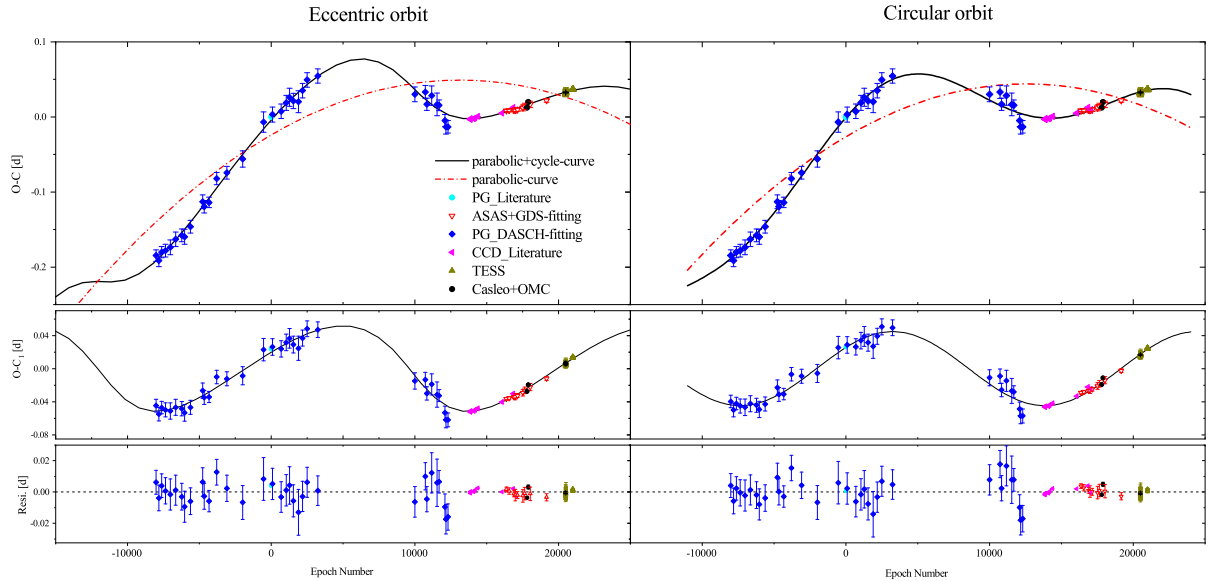}
\caption{The O-C diagrams of V606 Cen fitted with an eccentric orbit and a circular orbit, the data sources are listed in upper panel. The red dash lines are parabolic change. The black solid lines refer to downward parabolic plus cyclic variation in upper panel, and the cyclic variation alone is shown in the middel panel.}
\end{center}
\end{figure}

\begin{table}
\begin{center}
\caption{Orbital parameters of the third body in the V606 Cen system.}
\begin{tabular}{lll}\hline\hline
Parameters                                      &Eccentric orbit case                &Circular orbit case \\\hline
Revised epoch, $\Delta{T_{0}}$(d)               & $-0.02347\, (\pm0.00256)$               & $-0.02601\, (\pm0.0016)$\\
Revised period, $\Delta{P_{0}}$(d)              & $1.11\, (\pm0.01)\times{10^{-5}}$     & $1.13\, (\pm0.01)\times{10^{-5}}$\\
Semiamplitude, \,K(d)                               & $0.0545\, (\pm0.0046)$                & $0.0441\, (\pm0.0008)$\\
Orbital period, $P_{3}$ (yr)                   &$88.3\, (\pm1.9)$                     &$85.9\, (\pm1.3)$\\
Rate of the period change, $\dot{P}({d \,yr^{-1}})$  & $-2.08\, (\pm0.07)\times10^{-7}$ & $-2.22\, (\pm0.06)\times10^{-7}$\\
Longitude of the periastron passage, $\omega$(deg) &14.9\, ($\pm10.0$)                  &\nodata\\
Eccentricity, $e$                               & $0.33\, (\pm0.07)$                    &\nodata\\
Mass function, $f(m)$ ($M_{\odot}$)             & $0.126\, (\pm0.033)$                  & $0.060\, (\pm0.016)$ \\
Projected semi-major axis, $a_{12}\sin{i^{\prime}}$ (au) &$9.95\, (\pm0.85)$            &$7.64\, (\pm0.15)$ \\
Projected tertiary mass, $M_3\sin{i^{\prime}}$($M_{\odot}$) &$4.51\, (\pm0.43)$                &$3.42\, (\pm0.08)$\\
$\chi^{2}_{\nu}$                                 &$1.708\, (\pm0.018)$                    &$2.212\, (\pm0.034)$\\\hline
\end{tabular}
\end{center}
\end{table}

\section{Photometric Investigations with W-D Method}

The Wilson-Devinney (W-D) code used to model binary light curves has been constantly updated and improved since it came out in 1971 (Wilson \& Devinney 1971; Wilson 1990, 2012; Wilson \& Wyithe 2003). It has gradually become the most commonly used tool to analyze the light curves of binary stars. Up to now, only Lorenz et al. (1999) have analyzed the light curve of V606 Cen. They studied the light curve data from Mayer et al. (2010). The MORO code was used based on the W-D approach. Now we use the latest W-D version to analyze the TESS curve with higher accuracy. It is a continuous light curve, as shown in Fig. 2. We selected the data of tess-s0011-2-2 to study, because the observations are high quality and cover a continuous span of 27 days.

\begin{table}
\caption{Photometric solutions for V606 Cen}
\begin{center}
\setlength{\tabcolsep}{4mm}{
\begin{tabular}{llcc}\hline\hline
Parameters                                  &Solution H               &Complete data                &TESS data\\\hline
$\mathnormal{q}$(M$_{2}$/M$_{1}$)           &0.541\,($\pm0.001$)     &0.54845\,($\pm0.00074$)       &0.5263\,($\pm0.0025$)\\
T$_{1}$(K)                                  &29200\,(fixed)        &29000\,(fixed)        &29000\,(fixed)\\
T$_{2}$(K)                                  &21770\,($\pm20$)             &22073\,($\pm25$)             &22113\,($\pm16$)\\
$\mathnormal{i}$($^{\circ}$)                &87.3\,($\pm0.1$)         &88.30\,($\pm0.09$)        &86.44\,($\pm0.04$)\\
${\Delta}$T(K)                              &7430\,($\pm20$)          &6926\,($\pm25$)           &6886\,($\pm16$)\\
R$_{2}$/R$_{1}$                             &\nodata                       &0.75787\,($\pm0.00059$)     &0.7434\,($\pm0.0002$)\\
L$_{1}$/(L$_{1}$+L$_{2}$)($\mathnormal{U}$) &0.776\,($\pm0.002$)   &0.77911\,($\pm0.00095$)        &\nodata \\
L$_{2}$/(L$_{1}$+L$_{2}$)($\mathnormal{U}$) &0.224\,($\pm0.002$)   &0.22089\,($\pm0.00029$)         &\nodata \\
L$_{3}$/(L$_{1}$+L$_{2}$+L$_{3}$)($\mathnormal{U}$) &0.1\%\,($\pm0.05$)        &0.00\%\,($\pm0.14$)       &\nodata \\
L$_{1}$/(L$_{1}$+L$_{2}$)($\mathnormal{B}$) &0.758\,($\pm0.003$)   &0.75275\,($\pm0.00042$)          &\nodata \\
L$_{2}$/(L$_{1}$+L$_{2}$)($\mathnormal{B}$) &0.242\,($\pm0.003$)   &0.24725\,($\pm0.00042$)           &\nodata \\
L$_{3}$/(L$_{1}$+L$_{2}$+L$_{3}$)($\mathnormal{B}$) &1.0\%\,($\pm0.8$)    &0.34\%\,($\pm0.18$)       &\nodata \\
L$_{1}$/(L$_{1}$+L$_{2}$)($\mathnormal{V}$) &0.742\,($\pm0.002$)   &0.74532\,($\pm0.00038$)           &\nodata\\
L$_{2}$/(L$_{1}$+L$_{2}$)($\mathnormal{V}$) &0.257\,($\pm0.002$)   &0.25468\,($\pm0.00038$)            &\nodata \\
L$_{3}$/(L$_{1}$+L$_{2}$+L$_{3}$)($\mathnormal{V}$) & 0.2\%\,($\pm0.6$)  & 0.00\%\,($\pm0.16$)                   &\nodata \\
L$_{1}$/(L$_{1}$+L$_{2}$)($\mathnormal{TESS}$) &\nodata                        &\nodata                    &0.74156\,($\pm0.00012$)\\
L$_{2}$/(L$_{1}$+L$_{2}$)($\mathnormal{TESS}$) &\nodata                        &\nodata                    &0.25844\,($\pm0.00012$)\\
L$_{3}$/(L$_{1}$+L$_{2}$+L$_{3}$)($\mathnormal{TESS}$) &\nodata                &\nodata                  &1.775\%\,($\pm0.046$)\\
${\Omega}_{1}$=${\Omega}_{2}$               &2.942\,($\pm0.002$)     &2.96249\,($\pm0.00067$)     &2.92263\,($\pm0.00028$)\\
r$_{1}$(pole)                               &0.410\,($\pm0.001$)   &0.40750\,($\pm0.00094$)   &0.41057\,($\pm0.00031$)\\
r$_{1}$(side)                               &0.435\,($\pm0.001$)   &0.43219\,($\pm0.00029$)     &0.43571\,($\pm0.00041$)\\
r$_{1}$(back)                               &0.464\,($\pm0.001$)   & 0.46106\,($\pm0.00020$)    &0.46419\,($\pm0.00060$)\\
r$_{2}$(pole)                               &0.308\,($\pm0.001$)   &0.30811\,($\pm0.00029$)   &0.30440\,($\pm0.00012$)\\
r$_{2}$(side)                               &0.322\,($\pm0.001$)   &0.32193\,($\pm0.00036$)   &0.31791\,($\pm0.00014$)\\
r$_{2}$(back)                               &0.356\,($\pm0.001$)     &0.35490\,($\pm0.00061$)    &0.35069\,($\pm0.00024$)\\
the degree of contact($f$)     &0.04                &0.0165\,($\pm0.0021$)       &0.01095\,($\pm0.00089$)\\
Residual                                    &\nodata                    &0.00063              &0.00052\\
\hline
\end{tabular}}
\end{center}
\end{table}

Although the spectral type of V606 Cen was initially considered as B1-2 IB/II (Houk \& Cowley 1975), Lorenz et al. thought that it should be B0-0.5 V.
Based on V=9.887 mag for the primary star alone and E(B-V)=0.51 (Lorenz et al. 1999) and the distance obtained from the Gaia EDR3 (Gaia Collaboration 2020; Zari et al. 2021), we can estimate the absolute magnitude using the equation $M_{V} = m_{V}-5log(1000/\pi)+5-A_{V}$, where $\pi$ is the parallax in units of mas. We found an absolute magnitude for the primary of $M_{V}$ = -3.47 mag. This is just as expected for a B0.5 V star as listed on the online table\footnote{http$://$www.pas.rochester.edu/$\sim$emamajek/EEM$\_$dwarf$\_$UBVIJHK$\_$colors$\_$Teff.txt} by Mamajek (based mainly on Table 5 from Pecaut \& Mamajek 2013).
Therefore, the temperature of the primary star is set to be T = 29000\, K, which is similar to the temperature of solution H from Lorenz et al. (1999). The parameters for their solution H are listed in column 2 of Table 4. Next we performed our own fit of the photometry from Mayer et al. (2010) that was analyzed by Lorenz et al. However, instead of rebinning the data into averages of 11 to 22 individual measurements (as done by Lorenz et al.), we used all of the individual points in the fit. The results of this fit (column 3 of Table 4) are the same within errors as the solution H values. This indcates that time averaging (similar to the TESS cadence with 30 minute sampling) has no influence on the fit.

The space TESS project data can determine more reliable fundamental parameters by photometric solution.
The specific settings in W-D calculation are as follows: the gravity-darkening coefficients g$_{1}$ = g$_{2}$ = 1.0 (Lucy 1979) and the bolometric albedos A$_{1}$ = A$_{2}$ = 1.0 (Ruci{\'n}ski 1969) were adopted.
The specific intensities for the TESS wavelength band (dependent on local temperature, gravity, and viewing angle) were kindly provided to us from Professor Van Hamme. The contact model (model 3) was selected. The free parameters includes orbital inclination $\mathnormal{i}$, modified dimensionless surface potential of star 1 ${\Omega}_{1}$ (${\Omega}_{1}$ = ${\Omega}_{2}$ for mode 3), effective surface temperature of star 2 (T$_{2}$), bandpass luminosity of star 1 (L$_{1}$) and the mass ratio $\mathnormal{q}$. T$_{1}$ = 29000\, K was fixed, and the initial estimate of the mass ratio, 0.527, was adopted from the radial velocity measurements of Lorenz et al. (1999). First, we derived a convergent solution. Then, we tried to find solutions with other model configurations (W-D models 2, 4, and 5), but these did not converge to an acceptable solution. Therefore, we adopted the contact binary geometry for V606 Cen. The next step was to add the light of the third body, because of evidence of motion about a tertiary presented in Section 3. The flux contribution of the tertiary star $l_3$ was an adjustable parameter used in the final results. The corresponding photometric parameters are listed in Table 4. The theoretical light curves (solid line) are shown in Fig. 5 (TESS) and Fig. 6 (UBV from Mayer et al. 2010). The geometrical appearance of V606 Cen at orbital phases 0.0 and 0.25 is displayed in Fig. 7.

The second column of Table 4 is the solution H of Lorenz et al. (1999), the third column is the result by using the complete unbinned data of Mayer et al. (2010) with W-D code, and the fourth column gives the results from the TESS data. The radii are given as fractions of the semimajor axis, $R_1$ and $R_2$ are the equivalent volume radii, and the degree of contact is $f$ = ($\Omega_{in}$-$\Omega_{star}$)/($\Omega_{in}-\Omega_{out}$), where $\Omega_{star}$ is the modified dimensionless potential of the star surface, $\Omega_{in}$ and $\Omega_{out}$ are the dimensionless potential of the inner and outer Roche lobe, respectively.
All the solutions suggest that V606 Cen is a total-eclipsing shallow-contact binary ($f$ is less than 2$\%$). It is an A-subtype (the larger star is the hotter one), and the third light $\mathnormal{l}_3$ is less than 2$\%$ of the monochromatic flux. The results also reveal that the ${\Delta}$T is lower about 500\, K than the solution H, but the more massive component is about 7000\, K hotter than its less massive one. The mass ratio M$_{2}$/M$_{1}$ = 0.5263 of the TESS solution is closer to the mass ratio of $\mathnormal{q}_{spec}$ from the spectroscopic data.

\begin{figure}
\begin{center}
\includegraphics[angle=0,scale=0.4]{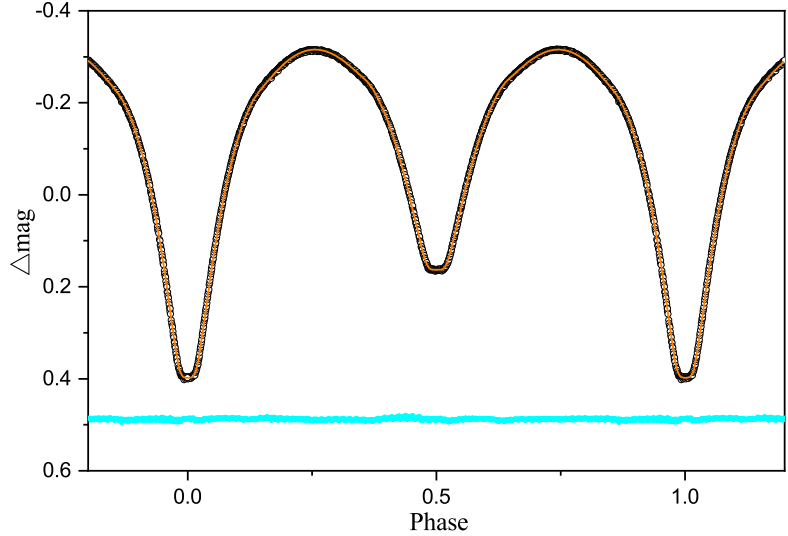}
\caption{The orange line and black open circles represent the theoretical and the observational light curve obtained by TESS, and Cyan open circles show the corresponding residuals}
\end{center}
\end{figure}

\begin{figure}
\begin{center}
\includegraphics[angle=0,scale=0.4]{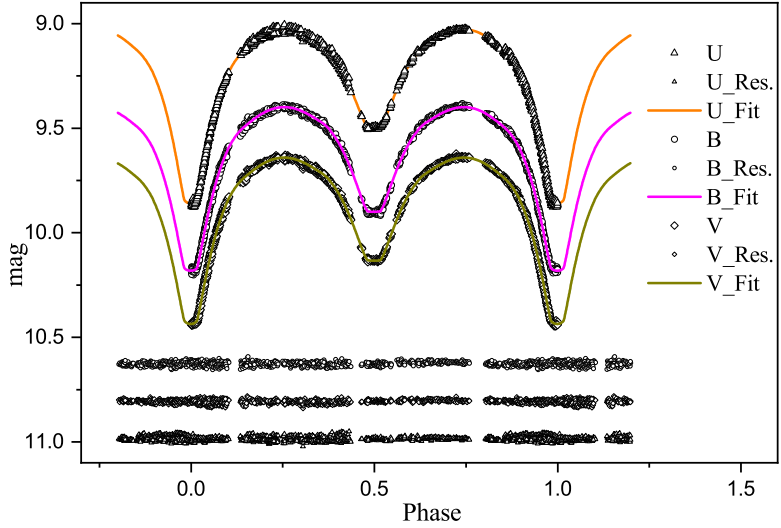}
\caption{The solid lines and open symbols represent the theoretical and observational  light curves obtained by Mayer et al., and below are their residuals (offset for $U$, $B$ and $V$ from top to bottom).}
\end{center}
\end{figure}

\begin{figure}
\begin{center}
\includegraphics[angle=0,scale=0.3]{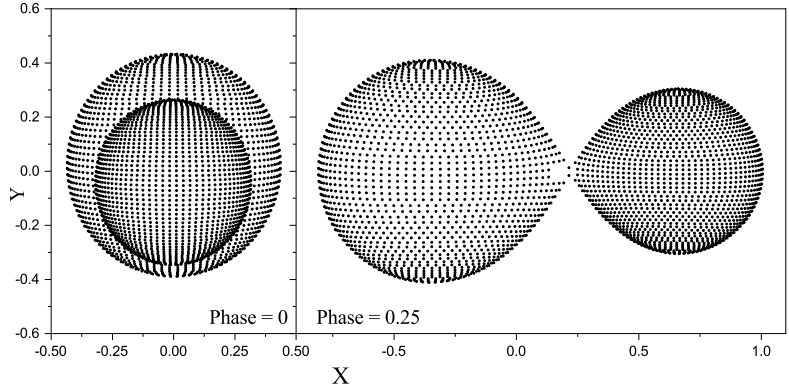}
\caption{The geometrical structure of V606 Cen for two orbital phases.}
\end{center}
\end{figure}

\section{Discussions and conclusions}

We have analyzed the periodic variation of V606 Cen for the first time based on our observations and a variety of online photometric databases. Its O-C diagram shows an obvious periodic oscillation, which is probably caused by the Light-Travel Time Effect due to the presence of the additional body (Liao \& Qian 2009; Qian et al. 2013a, 2013b; Zhao et al. 2019). We constructed and analyzed O-Cs in two cases: one is the eccentric orbit of the third body rotating around the central binary, a variation with a semi-amplitude of 0.0545\, d and a period of 88.3\, yr was detected. The other is a circular orbit solution with a semi-amplitude of 0.0441\, d and a period of 85.9\, yr. The value of $\chi^{2}_{\nu}$ for these two cases are 1.708 and 2.212. We can calculate the mass of third body using the equation (4):
\begin{equation}
{f}(m) = \frac{(M_{3}sini)^3}{(M_{1}+M_{2}+M_{3})^2} = \frac{4\pi^2}{GP_{3}^2}\times(a_{12}sini)^3.
\end{equation}
If we adopt binary masses of $M_1=14.7\,M_{\odot}$ (Lorenz et al. 1999), $M_2 = 7.74\,M_{\odot}$ (obtained by $M_{2}$/$M_{1}$), then the minimum mass of the third body is $M_3= 4.51(\pm0.43)\,M_{\odot}$ and $3.42(\pm0.08)\,M_{\odot}$ respectively for the eccentric and circular cases. If the orbital inclination of the third body is $\mathnormal{i} = 90^{\circ}$, the distance between the tertiary companion and binary system is $59.8(\pm6.4)$ and $58.5(\pm1.5)$ AU, respectively.

The photometric solutions imply that V606 Cen is a contact binary, and this binary just evolved to the contact stage with a fill-out factor of contact close to 0. The effective temperature of the more massive component is around 7000\, K hotter than that of its component. A high orbital inclination ($\mathnormal{i}$ = 86.44$^{\circ}$) was determined. Both the orbital inclination and the phased light curves indicate that this system is a totally eclipsing binary. Combining with the long-term orbital change shown in O-C diagram, the two cases show a downward parabolic change revealing a period decrease at a rate of $dP/dt = -2.08 \times{10^{-7}} d \cdot yr^{-1}$ and $dP/dt = -2.22 \times{10^{-7}} d \cdot yr^{-1}$ respectively for the eccentric and circular solutions. Considering the conservation of mass and angular momentum, the mean mass transfer rates can be yielded by the equation (5),
\begin{equation}
\dot{P}/P=3\dot{M}(1/M_{1}-1/M_2).
\end{equation}
$\dot{M} = +7.58\times{10^{-7}} M_{\odot}\cdot{yr^{-1}}$ and $+8.09\times{10^{-7}} M_{\odot}\cdot{yr^{-1}}$, respectively. The period decrease in V606 Cen can be explained by the mass transfer from the more massive component to the less massive one. This fact shows that the direction of mass transfer (primary to secondary) is the opposite of what Lorenz et al. expected for the evolutionary state of V606 Cen. For V606 Cen, its progenitor is a semi-detached binary. The less massive component continues to gain material from its companion, and a new contact binary star is born with both stars filling their Roche lobes.

The absolute magnitude of the secondary is $M_{V}$ = -2.53 mag by using the secondary to primary $V$-band flux ratio (Lorenz et al. 1999). Based on the table provided by Mamajek, we estimate the absolute magnitude of the secondary star from $T_{2}$ and $M_{2}$, $M_{V}$ is -2.11 and -1.93 mag. This indicates that the secondary component is overluminous for its mass and temperature. The reason may be that the higher temperature material was transferred from the primary star to the secondary component, and the secondary star obtains part of the energy, which makes the luminosity of the secondary component higher than a normal main sequence.

We can estimate the absolute magnitude of the tertiary from the minimum mass and the main sequence relations from Mamajek, $-0.71$ and $-0.03$ mag for the eccentric and circular orbits, respectively. This implies that the tertiary has a flux contribution of 5.2\% and 2.9\% in V-band flux for the eccentric and circular cases, respectively. Those are all larger than the third light (less than 2\%) estimated from the light curve. Thus, the light curve analysis appears to rule out any main sequence tertiary this bright. Is the tertiary body a compact body or a black hole? Further observations (e.g., the maximum angular separation, X-ray, etc.) might help establish whether or not the tertiary is a black hole.

Table 1 lists most of the known early OB-type contact binaries that have been investigated in detail. There are 18 contact binaries, 14 of which are located within the Milky Way and four beyond the Galaxy. There are nine objects whose orbital periodic changes have been studied, including one in an extragalactic target. However, only one of all the periodic variations of these binaries, GU Mon, is similar to that of V606 Cen. Both have the downward parabolic trend superimposed on a periodic variation. Interestingly, these two binaries are completely different in terms of their mass ratio and fill-out factor, i.e., the mass ratio is about 1 and 0.5, and the degree of contact is about 100\% and 0\%, respectively. Meanwhile, the contact degree of V606 Cen is the smallest in these binaries. We also discovered that the difference in effective temperature between the two components of V606 Cen is the largest except for SV Cen from the spectral type. All of the evidence shows that V606 Cen is in a special evolution stage, which provides a valuable test of the evolutionary models of the early OB-type binaries.

According to Kepler's third law and the parameters of photometric solution, the absolute parameters, such as the semi-major axes $a =15.53\,R_{\odot}$, $R_1=6.76\,R_{\odot}$, $R_2=5.05\,R_{\odot}$, are determined for V606 Cen. It is obvious that the component stars are very close to each other. This raises the question of how the binary could have avoided contact during the pre-main sequence stage when their radii were much larger. The additional body may be the key to this mystery. Due to the existence of the third body, the two stars with a relatively distant were brought together and evolved into an interacting binary (Naoz 2016; Qian et al. 2013c, 2014; Zhao et al. 2021; Eggleton 2012; Wang et al. 2021).

The formation and evolution of massive contact binaries is still poorly understood. There is some theoretical support (Sybesma 1985, 1986; Pols 1994; Qian et al. 2013a) that these early contact binaries may be formed through Case A mass transfer and the contact state is only a short-lived stage during the mass transfer. Some observations are in agreement with the prediction of the theories, such as BH Cen, V382 Cyg, TU Mus listed in Table 1. They are on the slow phase of Case A mass transfer with increasing periods. V606 Cen is in a unique stage of evolution different from that of these systems, and it is a newly formed massive contact binary created by mass transfer from the primary to the secondary. If its orbital period keeps decreasing, it will form a merger by common envelope evolution (Langer 2012), perhaps creating the kind of rapidly rotating object that may be a progenitor of gamma-ray burst during its ultimate collapse.

\acknowledgments
The authors thank the teams of DASCH database, ASAS and GDS survey for providing open data. This work is partly supported by the Chinese Natural Science Foundation (Nos. 11933008, 11873017 and 11903076) and the Yunnan Natural Science Foundation (No. 202001AT070051), and it is also partly supported by Science popularization department of Guizhou Association for Science and Technology and by Chinese Research Entity of Astronomical Technology and Education (CREATE). New photometric CCD observations were obtained at the Complejo Astronomico El Leoncito (CASLEO), San Juan, Argentina with a 0.60 m telescope. This paper includes data collected by the TESS mission. Funding for the TESS mission is provided by the NASA's Science Mission Directorate. Finally, we thank the referee for providing so many helpful suggestions.

%\bibliography{references}

\end{document}